\documentstyle[prd,aps,preprint]{revtex}
\tightenlines
\input epsf.tex
\def\DESepsf(#1 width #2){\epsfxsize=#2 \epsfbox{#1}}
\begin{document}

\draft
%\twocolumn[\hsize\textwidth\columnwidth\hsize\csname
%@twocolumnfalse\endcsname
\preprint{\vbox{
\hbox{UMD-PP-00-014} }}
\title{ Theories of Neutrino Masses and Mixings}
\author{ R. N. Mohapatra\footnote{e-mail:rmohapat@physics.umd.edu}}

\address{ $Department of
Physics, University of Maryland, College Park, MD 20742, USA$}
\date{October, 1999}
\maketitle
\begin{abstract}
Recent developments in our understanding of neutrino masses 
and their implications for physics beyond the standard model are
reviewed.

\end{abstract} 

\vskip0.5in

\section{Introduction}

History of weak interaction physics has to a large extent been a history
of our understanding of the properties of the elusive spin half particles
called the neutrinos. Evidence for only lefthanded neutrinos being emitted
in beta decay was the cornerstone of the successful V-A theory of weak
interactions suggested by Sudarshan, Marshak, Feynman and Gell-Mann; 
evidence for the neutral current interactions in early seventies 
provided brilliant confirmation of the successful gauge unification of
weak and electromagnetic interactions proposed by Glashow, Salam and
Weinberg.

Today as we enter a new millenium, we again have evidence for a very
important new property of the neutrinos i.e. they have mass and as a 
result
like the quarks, they mix with each other and lead to the phenomenon of
neutrino oscillation. This
is contrary to the expectations based on the standard model as well as
the old V-A theory (in fact one may recall that one way to make the V-A
theory plausible
was to use invariance of the weak Lagrangian under the so called
$\gamma_5$ invariance of all fermions, a principle which was motivated by
the assumption that
neutrinos have zero mass). The simple fact that neutrino masses vanish in
the standard model is proof that its nonzero mass is an indication of new
physics at some higher scale (or shorter distances). Study of details of
neutrino masses and mixings
is therefore going to open up new vistas in our journey towards a deeper
understanding of the properties of the weak interactions at very short
distances. This no doubt will have profound implications for the nature of
the final theory of the particles, forces and the universe. 

We are of course far from a complete picture of the masses and mixings of
the various neutrinos and cannot therefore have a full outline of the
theory that explains them. However there exist enough information and
indirect
indications that constrain the masses and mixings among the neutrinos that
we can see a narrowing of the possibilities for the theories. Many clever
experiments now under way will soon clarify or rule out many of the
allowed models. It will be one of the goals of this article to give a
panoramic view of the most likely scenarios for new physics that explain 
what is now known about neutrino masses\cite{book}. We hope to emphasize
the several interesting ideas for understanding the
small neutrino masses and discuss in general terms how they can lead to
scenarios for neutrinos currently being discussed in order to understand
the observations. These ideas have a very good
chance of being part of the final theory of neutrino masses. We then touch
briefly
on some specific models that are based on the above general framework
but attempt to provide an understanding of the detailed mass and mixing
patterns. These
works are instructive for several reasons: first they provide proof of
the detailed workability of the general ideas described above (sort of
existence proofs that things will work); second they often illustrate
the kind of assumptions needed and through that a unique insight into
which
directions the next step should be; finally of course nature may be 
generous in picking one of those models as the final message bearer.

\section{Neutrino mass is different from electron mass}

The electron and the neutrino are in many ways very similar particles:
they are both spin half objects; they both participate in weak
interactions with same
strength; in fact they are so similar that in the limit of exact gauge
symmetries they are two states of the same object and therefore in
principle indistinguishable. Yet there are profound differences between
them in the standard model: after gauge symmetry breaking, only
electron has electric charge. Another difference is that
only the lefthanded neutrino is included in the standard
model and not its righthanded counterpart; this is unlike the electron
both whose helicity states are included. In fact it is this property of
the standard model coupled with
$B-L$ being an exact symmetry which leads to neutrino
remaining massless to all orders in perturbation theory as well as
after the inclusion of nonperturbative effects.

The fact that the neutrino has no electric charge endows it with certain
properties not shared by other fermions of the standard model such
as the quarks and the electron (all of whom are electrically charged). The
point is that for neutral fermions one can write two kinds of Lorentz
invariant mass terms, the Dirac and Majorana masses, whereas for the
charged fermions, conservation of electric charge allows only the Dirac
type mass terms. In the four component notation for the fermions, the
Dirac
mass has the form $\bar{\psi}\psi$, whereas the Majorana mass is of the
form $\psi^TC^{-1}\psi$, where $\psi$ is the four component spinor and
$C$ is the charge conjugation matrix. One can also discuss the two
different kinds of mass terms using the two component notation for the
spinors. If we denote by $\chi$ and $\phi$ the two two-component spinors
which make up the four component object $\psi$, then
a Dirac mass is $\chi^T\sigma_2\phi$ whereas a Majorana mass is given by
$\chi^T\sigma_2 \chi$, where $\sigma_a$ are the Pauli matrices. To make
correspondence with the four component notation, we point out that $\chi$
and $i\sigma_2\phi^*$ are nothing but the $\psi_L$ and $\psi_R$
respectively. It is then clear that $\chi$ and $\phi$ have opposite
electric charges; therefore the Dirac mass $\chi^T\sigma_2\phi$ maintains
electric charge conservation (as well as any other kind of charge like
lepton number etc.).

This richness in the possibility for neutrino masses also has a down side
in the sense that in general, there are more parameters describing the
masses of the neutrinos than those for the quarks and leptons. For
instance 
for the electron and quarks, dynamics (electric charge conservation) 
reduces the number of parameters in their mass matrix. As an example,
using the two component notation for all fermions, for the case of two
two component spinors, a charged fermion mass will be described only by
one parameters whereas for a neutrino, there will be three parameters.
This difference inceases rapidly e.g. for 2N spinors, to describe charged 
fermion masses, we need $N^2$ parameters (ignoring CP violation) whereas
for neutrinos, we need $\frac{2N(2N+1)}{2}$ parameters. What is more
interesting is that for a neutrino like particle, one can have both even
and odd number of two component objects and have a consistent theory.
There are suggestions that if there are an odd number of two component 
charged spinors, the corresponding theory is not consistent as a
perturbative field theory.

In this article, we will use two component notation for neutrinos. Thus
when we say that there are N neutrinos, we will mean N two-component
neutrinos. 

In the two component language, all massive neutrinos are Majorana
particles and what is conventionally called a Dirac neutrino is really a
very specific choice of mass parameters for the Majorana neutrino. 
Let us give some examples: If there is only one two component neutrino (we
will drop the prefix two component henceforth), it can have a mass
$m\nu^T\sigma_2\nu$(to be called $\equiv m\nu\nu$ in shorthanded
notation). The neutrino is now a self conjugate object which can be seen
if we write an equivalent 4-component spinor $\psi$:
\begin{eqnarray}
\psi=\left(\begin{array}{c} \nu \\ i\sigma_2\nu^* \end{array}\right)
\end{eqnarray}
Note that this 4-component spinor satisfies the condition 
\begin{eqnarray}
\psi~=~\psi^c\equiv C\bar{\psi}^T
\end{eqnarray}
This condition implies that the neutrino is its own anti-particle, a
 fact more transparent in the 4- rather than the two-component
notation. The above exercise illustrates an important point i.e. given
any two component spinor, one can always write a self conjugate (or
Majorana)
4-component spinor. Whether a particle is really its own antiparticle or
not is therefore determined by its interactions. To see this for the
electrons, one may solve the following excercise i.e. if we wrote two
Majorana spinors using the two two component spinors that describe the
charged
fermion (electron), then until we turn on the electromagnetic
interactions and the mass term, we will not know whether the electron is
its own antiparticle or not. Once we turn on the electromagnetism,
this ambiguity is resolved.

Let us now go one step further and consider two 2-component neutrinos
($\nu_1,~\nu_2$). The general mass matrix for this case is given by:
\begin{eqnarray}
{\cal M}_{2\times 2}~=~\left(\begin{array}{cc} m_1 & m_3 \\ m_3 & m_2
\end{array}\right)
\end{eqnarray}
Note first that this is a symmetric matrix and can be diagonalized by
orthogonal transformations. The eigenstates which will be certain
admixtures of the original neutrinos now describe self conjugate
particles. One can look at some special cases:

\noindent {\it \underline{Case i}:} 

If we have $m_{1,2}=0$ and $m_3\neq 0$, then one can 
assign a charge +1 to $\nu_1$ and -1 to $\nu_2$ and the theory has an
extra $U(1)$ symmetry which can be identified as the lepton number and the
particle is then called a Dirac neutrino. The point to be noted is that
the Dirac neutrino is a special case of for two Majorana neutrinos.
In fact if we insisted on calling this case one with two Majorana
neutrinos, then the two will have equal and opposite (in sign) mass as can
be seen
diagonalizing the above mass matrix. Thus a Dirac neutrino can be thought
of as two Majorana neutrinos with equal and opposite (in sign) masses.
Since the argument of a complex mass term in general refers to its C
transformation property (i.e. $\psi^c = e^{i\delta_m}\psi$, where
$\delta_m$ is the phase of the complex mass term), the two two component
fields of a Dirac neutrino have opposite charge conjugation properties.

\noindent {\it \underline{Case ii}:} 

If we have $m_{1,2}\ll m_3$, this case is called
pseudo-Dirac neutrino since this is a slight departure from case (i). In
reality, in this case also the neutrinos are Majorana neutrinos with their
masses $\pm m_0 +\delta$ with $\delta \ll m_0$. The two component
neutrinos will be maximally mixed. Thus this case is of great current
physical interest in view of the atmospheric (and perhaps solar) neutrino
data.

\noindent {\it \underline{ Case iii}:}

 There is third case where one may have $m_1=0$
and $m_3\ll m_2$. In this case the eigenvalues of the neutrino mass matrix
are given respectively by: $m_{\nu}\simeq -\frac{m^2_3}{m_2}$ and $M\simeq
m_2$.
One may wonder under what conditions such a situation may arise in
a realistic gauge model. It turns out that if $\nu_1$ transforms as an
$SU(2)_L$ doublet
and $\nu_2$ is an $SU(2)_L$ singlet, then the value of $m_3$ is limited by
the weak scale whereas $m_2$ has no suchlimit and $m_1=0$ if the theory
has no
$SU(2)_L$ triplet field (as for instance is the case in
the standard model). Choosing $m_2\gg m_3$ then provides a
natural way to understand the
smallness of the neutrino masses. This is known as the seesaw mechanism
\cite{grsyms}. Since this case is very different from the case (i) and
(ii), it is generally said that in grand unified theories, one expects the
neutrinos to be Majorana particles. The reason is that in most grand
unified theories there is a higher scale which under appropriate
situations provides a natural home for the large mass $m_2$. 

While we have so far used only two
neutrinos to exemplify the various cases including the seesaw mechanism,
these discussions generalize when $m_{1,2,3}$ are each $N\times N$
matrices (which we denote by $M_{1,2,3}$). For example, the seesaw formula
for this general situation can
be written as
\begin{eqnarray}
{\cal M}_{\nu} \simeq - M^T_{3D}M^{-1}_{2R} M_{3D}
\end{eqnarray}
where the subscripts $D$ and $R$ are used in anticipation of their origin
in gauge theories where $M_D$ turns out to be the Dirac matrix and $M_R$
is the mass matrix of the right handed neutrinos and all eigenvalues of
$M_R$ are much larger than the elements of $M_D$.
It is also worth pointing out that
Eq. (4) can be written in a more general form where the
Dirac matrices are
not necessarily square matrices but $N\times M$ matrices with $N\neq
M$. We give such examples below.

 Although there is no experimental proof
that the neutrino is a Majorana particle, the general opinion is that the
the seesaw mechanism provides such a simple way to understand the glaring
differences between the masses of the neutrinos and the charged fermions
that neutrinos indeed must be Majorana particles.

Even though for most situations, the neutrino can be treated as a two
component object regardless of whether its mass is of Dirac or Majorana
type, there are certain practical situations where differences between the
Majorana and Dirac neutrino becomes explicit: one case is when the two
neutrinos annihilate. For Dirac neutrinos, the particle and the
antiparticle are distinct and therefore there annihilation is not
restricted by Pauli principle in any manner. However, for the case of
Majorana neutrinos, the identity of neutrinos and antineutrinos plays an
important role and one finds that the annihilation to the Z-bosons occurs
only via the P-waves. Similarly in the decay of the neutrino to any final
state, the decay rate for the Majorana neutrino is a factor of two
higher than for the Dirac neutrino. 

\section{Experimental indications for neutrino masses}

As has been extensively discussed elsewhere in this book, while the direct
search experiments for neutrino masses using tritium beta decay and
neutrinoless double beta decay have only yielded upper limits, the
searches for neutrino oscillation, which can occur only if the neutrinos
have masses and mixings have yielded positive evidences. There
is now clear evidence from one and strong indications from other
experiments for neutrino oscillations and hence
neutrino masses. The evidence comes from the atmospheric neutrino data
in the Super-Kamiokande
experiment\cite{sk} which confirms the indications of oscillations in
earlier data from the Kamiokande\cite{kam}, 
IMB\cite{IMB} experiments. More recent data from Soudan II\cite{soudan}
and MACRO\cite{macro} experiments provide further confirmation of this
evidence. The observation here is the following: in the standard model
with massless neutrinos, all the muon and electron
neutrinos produced at the top of the atmosphere would be expected to
reach detectors on the earth and would be isotropic; what has been
observed is that while that is true for the electron neutrinos, the muon
neutrino flux observed on earth exhibit a strong zenith angle dependence.
A simple way to understand this would be to assume that the muon neutrinos
oscillate into another undetected species of neutrino on their way to the
earth, with a characteristic oscillation
length of order of ten thousand kilometers. Since the oscillation length
is roughly given by $E(GeV)/{\Delta m^2(eV^2)}$ kilometers, for a GeV
neutrino, one would expect
the particle physics parameter $\Delta m^2$ corresponding to the mass
difference between the two neutrinos to be around $10^{-3}$ eV$^2$
corresponding to maximal mixing.

From the existing data several important conclusions can be drawn:
(i) the data cannot be fit assuming oscillation between $\nu_{\mu}$ and
$\nu_e$; (ii) two oscillation scenarios that fit the data are
$\nu_{\mu}-\nu_{\tau}$ as well as $\nu_{\mu}-\nu_s$ oscillations
(where $\nu_s$ is a sterile neutrino that
does not couple to the W or the Z bosons in the
basic Lagrangian), although
at the two $\sigma$ level, the first scenario is a better fit than the
latter. The more precise values of the oscillation parameters
at 90\% c.l. are:
\begin{eqnarray}
\Delta m^2_{\nu_{\mu}\nu_{\tau}}\simeq (2~-~8)\times 10^{-3}~ eV^2;\\
\nonumber
sin^2 2\theta_{\mu\tau}\simeq 0.8~-~1
\end{eqnarray}

The second evidence for neutrino oscillation comes from the five
experiments that have observed a deficit in the flux of neutrinos from the
Sun as compared to the predictions of the standard solar model championed
by Bahcall and his collaborators\cite{bahcall} and more recently studied
by many groups. The experiments resposible for this discovery are the
Chlorine, Kamiokande, Gallex, SAGE and Super-Kamiokande\cite{solar}
experiments conducted at the Homestake mine, Kamioka in Japan, Gran Sasso
in Italy and Baksan in Russia. The different experiments see different
parts of the solar neutrino spectrum. The details of these considerations
are discussed in other chapters. The oscillation
interpretation of the solar neutrino deficit has more facets to it than
the atmospheric case: first the final state particle that the $\nu_e$
oscillates into and second what kind of $\Delta m^2$ and mixings fit the
data. At the moment there is a multitude of possibilities. Let us
summarize them now. 

As far as the final state goes, it can either be one of the two remaining
active neutrinos , $\nu_\mu$ and $\nu_{\tau}$ or it can be the sterile
neutrino $\nu_s$ as in the case of atmospheric neutrinos. Both
possibilities are open at the moment. The SNO
experiment which is expected to measure the solar neutrino flux via 
neutral current interactions will settle the issue of whether the final
state of solar neutrino oscillation is the active neutrino or the sterile
neutrino; in the former case, the ratio $r$ of charged current flux
($\Phi_{CC}$) to the neutral current flux ($\Phi_{NC}$) is nearly half
whereas in the latter case, it is one. As far as the $\Delta m^2$ and
$sin^22\theta$ parameters go, there are three possibilities: if the
oscillation proceeds without any help from the matter in the dense core
of the Sun, it is called vacuum oscillation (VO); if oscillation is
enhanced by the solar core, it is called the MSW mechanism, in which case
there are two ranges of parameters that can explain the deficit- the small
angle (SMA-MSW) and the large angle (LMA-MSW). The parameter ranges, taken
from Ref.\cite{bahcall} are:
\begin{eqnarray}
VO:  \Delta m^2 \simeq 6.5\times 10^{-11} eV^2; sin^22\theta\simeq 0.75 
-1 \\ \nonumber
SMA-MSW: \Delta m^2 \simeq 5\times 10^{-6} eV^2;  sin^22\theta \simeq
5\times 10^{-3} \\ \nonumber
LMA-MSW:  \Delta m^2 \simeq 1.2\times 10^{-5}-3.1\times 10^{-4} eV^2;
sin^22\theta \simeq 0.58~-~1.00
\end{eqnarray}
Relevant point to note is that there is no large angle MSW fit for the
case of sterile neutrinos due to absence of matter effect for them. The
situation lately however has been quite fluid in the sense that there are
measurements from the Super-Kamiokande experiment of the electron energy
distribution which appears to contradict the MSW-SMA solution; similarly
there are day night effects which seem to prefer MSW-LMA solution although
VO solutions also lead to day night effects due to matter effects for
certain ranges of $\Delta m^2$\cite{guth}. There are also indications of 
seasonal variation of the solar neutrino flux above and beyond that
expected from the position of the earth in the orbit.

Finally, we come to the last indication of neutrino oscillation from the 
Los Alamos Liquid Scintillation Detector (LSND) experiment\cite{lsnd} ,
where neutrino oscillations of $\nu_{\mu}$ both from a stopped muon decay 
(DAR) as well as the
one accompanying the muon in pion decay
(DIF) have been observed. The evidence from the DAR is
statistically
more significant and is an oscillation from $\bar{\nu}_\mu$ to
$\bar{\nu}_e$. The mass and mixing parameter range that fits data is:
\begin{eqnarray}
LSND: \Delta m^2 \simeq 0.2 - 2 eV^2; sin^22\theta \simeq 0.003-0.03
\end{eqnarray}
There are also points at higher masses specifically at 6 eV$^2$ which are
 allowed by the present LSND data for small mixings\cite{louis}. 
Presently KARMEN experiment at the Rutherford laboratory is also searching
for
the same oscillation. While they have found about eight events as of this
writing, this is consistent with their expected background\cite{drexlin}.
The proposed MiniBooNE experiment Fermilab\cite{louis} will provide more
definitive information on this very important process in the next five
years.

Our goal now is to study the theoretical implications of these
discoveries. We will proceed towards this goal in the following manner: 
we will isolate the mass patterns that fit the above data and then
look for plausible models that can first lead to the general 
feature that neutrinos have tiny masses; then we would try to understand
in simple manner some of the features indicated by data in the hope 
that these general ideas will be part of our final understanding of the
neutrino masses. As mentioned earlier on, to understand the neutrino
masses one has to go
beyond the standard model. First we will sharpen what we mean by this
statement. Then we will present some ideas which may form the basic
framework for constructing the detailed models.
We will refrain from discussing any specific models
except perhaps giving examples by way of illustrating the theoretical
ideas.

\section{Patterns and textures for neutrinos}

As already mentioned, we will assume two
component neutrinos and therefore their masses will in general be Majorana
type. Let us also give our notation to facilitate further discussion: 
the neutrinos emitted in weak processes such as the beta decay or muon 
decay are weak eigenstates and are not mass eigenstates. The latter
determines how a neutrino state evolves in time. Similarly, in the
detection process, it is the weak eigenstate that is picked out. This is
of course the key idea behind neutrino oscillation\cite{maki}. It is
therefore important to express the weak eigenstates in terms of the mass
eigenstates. We will denote
the weak eigenstate by the symbol $\alpha, \beta$ or simply $e,\mu, \tau$ 
etc whereas the mass
eigenstate will be denoted by the symbols $i,j,k$ etc. The
mixing angles will be 
denoted by $U_{\alpha i}$ and relate the two sets of eigenstates as
follows:
\begin{eqnarray}
\left(\begin{array}{c} \nu_e\\ \nu_{\mu} \\ \nu_{\tau}\end{array}
\right)=~ U\left(\begin{array}{c} \nu_1 \\ \nu_2 \\ \nu_3
\end{array} \right)
\end{eqnarray}
Using this equation, one can derive the wellknown oscillation formulae for
the survival probability of a particular weak eigenstate $\alpha$ to be:
\begin{eqnarray}
P_{\alpha\alpha}= 1-4\sum_{i< j} |U_{\alpha i}|^2 |U_{\alpha j}|^2 sin^2
\Delta_{ij}
\end{eqnarray}
where $\Delta_{ij}=\frac{(m^2_i-m^2_j)L}{4E}$. The transition probability
from one weak eigenstate to another is given by
\begin{eqnarray}
P_{\alpha\beta}=~4\sum_{i<j} U_{\alpha i} U^*_{\beta i} U^*_{\alpha j}
U_{\beta j} sin^2\Delta_{ij}
\end{eqnarray}

Since Majorana masses violate lepton number, a very important
constraint on any discussion of neutrino mass patterns arises from the
negative searches for neutrinoless double beta decay\cite{klapdor}. 
The most stringent present limits are obtained from the Heidelberg-Moscow
enriched Germanium-76 experiment at Gran Sasso and implies an upper limit
on the following combination of masses and mixings:
\begin{eqnarray}
<m_{\nu}>\equiv \Sigma_i U^2_{ei} m_{\nu_i} \leq 0.2 ~eV
\end{eqnarray}
This upper limit depends on the nuclear matrix element calculated by the
Heidelberg group\cite{klap1}. There could be an uncertainty of a factor of
two to three in this estimate. This would then relax the above upper bound
to at most 0.6 eV. This is a very stringent limit and becomes especially
relevant when one considers whether the neutrinos constitute a significant
fraction of the hot dark matter of the universe. A useful working formula
is $\Sigma_i m_{\nu_i}\simeq 24\Omega_{\nu} $ eV where $\Omega_{\nu}$ is
the neutrino fraction that contributes to the dark matter of the universe.
For instance, if the dark matter fraction is 20\%, then the sum total of
neutrino masses must be 4.8 eV. The situation at the moment is
uncertain\cite{primack} after the results from the high z supernova
searches indicated possible nonzero cosmological constant. Nevertheless,
from structure formation data, a total neutrino mass of 2-3 eV cannot
strictly be ruled out and in fact one particular fit\cite{silk} prefers
a cold+hot dark matter with a similar mass as a better fit than any other
(e.g. CDM+$\Lambda$). The proposed GENIUS\cite{genius} experiment by the
Heidelberg-Moscpw collaboration has the promise to bring down the upper
limit on $<m_{\nu}>$ by two orders of magnitude. This would profoundly
effect the current ideas on neutrino masses and will help to more sharply
define the theoretical directions in the field. 

In view of several levels of uncertainties that currently surround the
various
pieces of information on neutrino masses, we will consider different
scenarios. It is not an unfair reflection of the present community
consensus  to say that the solar and the atmospheric neutrino results
are the most secure evidences for neutrino masses. We will therefore
first consider the implications of taking these two sets of data seriously
supplemented by the very useful information from neutrinoless double beta
decay. We will include the LSND data subsequently.

\subsection{Solar and Atmospheric data and neutrino mass patterns}

If one wants to fit only the solar and atmospheric neutrino data,
regardless of the nature of the solution to the solar neutrino puzzle
(i.e. MSW small
or large angle or vacuum oscillation), it is possible to have consistent
scenarios
using only the three known neutrinos. There are many mixing patterns and
neutrino mass matrices that can be used for the purpose. In discussing
these patterns, it is important to remember that a solution to the
atmospheric neutrino puzzle requires that in the context of a three
neutrino model, $\nu_{\mu}$ and $\nu_{\tau}$ must mix maximally. 
There are two interesting mass patterns that have been widely discussed in
literature: (i) hierarchical pattern with $m_1\ll m_2 \ll m_3$ or 
(ii) approximately degenerate pattern \cite{cald1} $m_1\simeq m_2 \simeq
m_3$, where $m_i$ are the eigenvalues of the neutrino mass matrix. In
the first case, the atmospheric and the solar neutrino data give direct
information on $m_3$ and $m_2$ respectively. On the other hand,
in the second case, the mass differences between the first and
the second eigenvalues are chosen to fit the solar neutrino data and
the second and the third to fit the atmospheric neutrino data.
Neutrinoless double beta decay limits imply very stringent constraints on
the mixing
pattern in the degenerate case; but before we proceed to this discussion
let us focus for a while on the hierarchical mass pattern. 

In proceeding with this discussion it is important to remember that the
CHOOZ reactor data\cite{chooz} implies that for $\Delta m^2\geq 10^{-3}$
eV$^2$, the
electron neutrino mixing angle has a rough upper bound $|U_{ei}|\leq
0.2$. Furthermore it is now certain that atmospheric neutrino data
cannot be fitted by $\nu_{\mu}-\nu_e$ oscillation. Together
they imply that in any neutrino mass matrix construction, one
must require that  $|U_{e3}|\leq 0.2$\cite{lisi}. Note that $U_{e2}$ can
be larger since the relevant $\Delta m^2$ which corresponds to the solar
neutrino puzzle is lower than what the CHOOZ experiment is sensitive to. 
On the other hand the
solar neutrino admits both small and large mixing angles. Combining these
two inputs, one can conclude that the $3\times 3$ neutrino mixing matrix
is given by\cite{gold}
\begin{eqnarray}
U_{\nu}=\left(\begin{array}{ccc}
c & s & 0\\
\frac{s}{\sqrt{2}} &\frac{c}{\sqrt{2}} & -\frac{1}{\sqrt{2}} \\
\frac{s}{\sqrt{2}} &\frac{c}{\sqrt{2}} &+\frac{1}{\sqrt{2}}
\end{array}\right)
\end{eqnarray}
In writing the above mixing matrix, we have assumed atmospheric
neutrino data fit with $sin^22\theta =1$. On the other hand if we took,
this value to be $sin^22\theta =8/9$, the above mixing matrix changes to:
\begin{eqnarray}
U_{\nu}=\left(\begin{array}{ccc}
c & s & 0\\
\frac{s}{\sqrt{3}} &\frac{c}{\sqrt{3}} & -\frac{\sqrt{2}}{\sqrt{3}} \\
\frac{s\sqrt{2}}{\sqrt{3}} &\frac{c\sqrt{2}}{\sqrt{3}}
&+\frac{1}{\sqrt{3}}
\end{array}\right)
\end{eqnarray}
There are two special cases, where these mixing matrices take specially 
appealing forms: case (A): if maximal mixing solutions are chosen for the
solar neutrino puzzle or case(B): it is assumed that the neutrino masses
are degenerate with a mass $m_0$ such that the effective mass deduced from
neutrinoless double beta decay $<m_{\nu}> \ll m_0$\cite{vissani}. In
either case, the first matrix reduces to the so called bimaximal
form\cite{gold,vissani,feru} whereas the second matrix reduces to the
democratic
form\cite{xing}. The mixing matrices are obtained from the above two
equations by taking $c=s=\frac{1}{\sqrt{2}}$. The neutrino mass matrices
in both these cases can be obtained by writing
\begin{eqnarray}
{\cal M_{\nu}}= U M^d U^T
\end{eqnarray}
where $M^d = Diag(m_1,~m_2,~m_3)$.

\bigskip

\noindent{\bf \underline{Degenerate case}:}

\bigskip

The most compelling physical motivation for this case comes from the
requirement that neutrinos
constitute a significant fraction of the dark matter of the universe.
Using our previous discussion, we see that if the neutrino HDM constitutes
about 20\% of the critical mass of the universe, then the total neutrino
mass i.e. $\Sigma_i |m_i|$ must be about 4.8 eV. If all three neutrinos
are degenerate the share of each species is 1.6 eV. Note that this is much
bigger than the present upper limits on the effective mass $<m_{\nu}>$
from neutrinoless double beta decay. If we ignore CP phases, then this
implies a strong constraint on the mixings.

Two particularly interesting mass matrices emerge, for the case where the
neutrino masses are approximately degenerate. To derive them, let us
choose $M^d= Diag(m_0,~-m_0,~m_0)$. If we further assume that
$U_{e3}\simeq 0$, then we conclude that the mixing matrix elements
$U_{e1}\simeq U_{e2}\simeq \frac{1}{\sqrt{2}}$. It therefore follows that
the small angle MSW solution is automatically eliminated.
 For the bimaximal
case, one gets the neutrino mass matrix of the form\cite{georgi}:
\begin{eqnarray}
M_{\nu}= m_0\left(\begin{array}{ccc} 0 & \frac{1}{\sqrt{2}} &
\frac{1}{\sqrt{2}} \\ \frac{1}{\sqrt{2}} & \frac{1}{2} & -\frac{1}{2}\\
\frac{1}{\sqrt{2}} & -\frac{1}{2}  & \frac{1}{2} \end{array}\right)
\end{eqnarray}
For derivation of this mass matrix in gauge models, see Ref.\cite{nuss}.

There is a corresponding mass matrix for the democratic case: it is given
by
\begin{eqnarray}
M_{\nu}= m_0\left(\begin{array}{ccc} 0 & \frac{1}{\sqrt{3}} &
\frac{\sqrt{2}}{\sqrt{3}} \\ \frac{1}{\sqrt{3}} & \frac{2}{3} &
-\frac{\sqrt{2}}{3}\\
\frac{\sqrt{2}}{\sqrt{3}} & -\frac{\sqrt{2}}{3}  & \frac{1}{3}
\end{array}\right)
\end{eqnarray}

Finally we want to note the form of the neutrino mass matrices that lead
in general case to mixing matrices which have either the democratic or the
bimaximal form regardless of the nature of the eigenvalues. The importance
of this discussion is that in trying to construct gauge models for
neutrinos, the mass matrix follows directly from the Lagrangian and the
mixing follows from this afterwards. 

\noindent{\it (i) Bimaximal case}

\begin{eqnarray}
M_{\nu}=\left(\begin{array}{ccc} A+D & F & F \\ F & A & D \\ F & D & A
\end{array}\right)
\end{eqnarray}
Note that vanishing of neutrinoless double beta decay implies that $A=-D$.
A special case of this is when $A=D=0$. Such mass matrices have been
discussed in literature for the three neutrino case in Ref.\cite{barb} and
for construction of gauge models for this casee, see \cite{josh}.

\bigskip

\noindent{\it (ii) Democratic case:}

This case is more complicated. As was noted in Ref.\cite{nuss2}
one must choose the charged lepton mass matrix in the form while keeping
the neutrino mass matrix diagonal.:
\begin{eqnarray}
M_{\ell}=\left(\begin{array}{ccc} a & 1 & 1 \\ 1 & a & 1 \\ 1 & 1 & a
\end{array}\right) 
\end{eqnarray}
The symmetric form of the matrix in the democratic case is clear and in
fact it exhibits a permutation symmetry $S_3$ operating on the lepton
doublets, which provides another clue to possible gauge model building.

\bigskip
\noindent{\bf \underline{Hierarchical case}:}

\bigskip

One can also get some clue to model building by analysing the case where
the neutrino masses are hierarchical i.e.
$m_1\ll m_2 \ll m_3 $. To see this let us first ignore the couplings of
the first generation and consider only the $2\times 2$ mass matrix
involving the second and the third generation. It has been pointed out
that\cite{alt}, one very simple way to get a natural hierarchy while
having a maximal (or large) mixing is to have a matrix $M_{\nu}$ of rank
one i.e.
\begin{eqnarray}
M_{\nu}=\left(\begin{array}{cc} x^2 & x\\ x & 1 \end{array}\right)
\end{eqnarray}
This matrix has one zero eigenvalue whereas choosing $x\simeq 1$ leads to
large mixing angle. Further interest in this idea emerges
from the observation that such an $M_{\nu}$ can arise via the seesaw
mechanism if one chooses $M_D=\left(\begin{array}{cc} 0 & 0\\ x &
1\end{array}\right)$. The interesting point about this form for $M_D$ is
that if we define it as $\bar{\nu}_L M_{D}\nu_R$ (with $\nu_{L,R}$ are
column vectors), the above form for $M_D$ only mixes the right handed
neutrinos and not the left handed ones. In a quark lepton unified model
this would mean that only right handed quarks mix with a large mixing
angle. This of course is completely unconstrained by observations that
confirm the standard
model since there are no righthanded charged currents in the standard
model. More
importantly, it leaves the left handed mixing small. Thus this observation
may provide one reconciliation of the apparent conundrum that the quark
mixing angles are small whereas the neutrino mixng angles are large.
This particular form has the disadvantege that it cannot arise in models
such as left-right models which generally lead to symmetric Dirac masses.

There are also other ways to generate maximal mixing angles. For instance
a choice for neutrino mass matrix (for the 2-3 sector) of the form
\begin{eqnarray}
M_{23}=\left(\begin{array}{cc} a & b \\ b & a \end{array}\right)
\end{eqnarray}
leads to maximal mixing.
These mass matrices respect permutation symmetries (e.g. $S_3$) and
therefore can be stable under radiative corrections. Ref.\cite{akh} has
discussed the stability of
neutrino mass matrices with respect to small variations of the parameters
for various neutrino mass hierarchies.

\subsection{Solar, atmospheric and LSND data and scenarios with sterile
neutrinos}

In order to explain solar, atmospheric and the LSND data simultaneously
using the oscillation picture, one must invoke additional neutrinos since
the different $\Delta m^2$s needed to explain each piece of data is very
different from the others and would never add up to zero as would be the
case if there were only three neutrino species i.e. $\Delta m^2_{12} +
\Delta m^2_{23} +\Delta m^2_{31}=0$. This is a revolutionary result since
LEP and SLC measurements imply precisely (within very small errors)
that there are almost precisely three active neutrinos coupling to the
Z-boson (the actual LEP result is $N_{\nu}=2.994\pm 0.012$\cite{lep}).
Furthermore, even though the conclusions from big bang
nucleosynthesis\cite{bbn} are not as
precise as the LEP-SLC data, the Helium abundance is very sensitive to the
number of neutrinos and anywhere from 3.2 to 4.4 total number of neutrinos
could be accomodated. Already these
two results imply strong constraints on the theories that include extra
neutrinos beyond those present in the standard model. For instance, the
LEP-SLC data data implies that the extra neutrinos cannot couple to the
Z-boson with the same strength as other neutrinos. They
have therefore been designated in the literature as sterile neutrinos.
Their ``sterility'' also helps them to evade the bounds from big bang
nucleosynthesis as follows. Suppose, the coupling of a sterile neutrino to 
known leptons is given by the four Fermi interaction with a strength given
by $G_F\epsilon$. Then, if $\epsilon\leq 10^{-3}$\cite{book}, then the
sterile neutrinos decouple before the QCD phase transition temperature
of about a 100 MeV and their effective contribution at the BBN era becomes
equivalent to about $0.1$ neutrino species. There are further constraints
on $\Delta m^2$ and $sin^22\theta$
that arise when the sterile neutrinos oscillate at the BBN
era\cite{chizov}. Typically, the constraint is
\begin{eqnarray}
\Delta m^2_{es} sin^22\theta_{es} \leq 10^{-6} eV^2
\end{eqnarray}
 These considerations are
important in constructing detailed theories of the sterile neutrinos. 
While we postpone the detailed discussion of theories with extra
neutrinos to a separate section, here we wish to study the possible four
or more neutrino patterns and their mass matrices that correspond to the
exsting neutrino data.

\bigskip

\noindent{\it \underline{B1. Four neutrino case}}
\bigskip

It is possible to construct three possible scenarios for neutrino data
using the four neutrinos. Denoting the sterile neutrino by $\nu_s$, the
three scenarios are the following:

\noindent{\it \underline{Case B1.1:}} 

The solar neutrino data is solved via the MSW small
angle oscillation between the $\nu_e-\nu_s$, which then have a mass
difference of $\Delta m^2_{es}\simeq 10^{-5}$ eV$^2$ whereas the
atmospheric neutrino data is solved via the $\nu_{\mu}-\nu_{\tau}$
oscillation\cite{cald1,pelto}.  (Of course, the solar neutrino
puzzle could also be solved via the vacuum oscillation of $\nu_e-\nu_s$.)
In either case, if we have
$m_{\nu_{\mu}}\simeq m_{\nu_{\tau}}\simeq $ eV, then there can be a
significant hot dark matter component in the universe with enormous
implications for structure formation. Note that LSND data is then fitted
by $\nu_{\mu}-\nu_e$ oscillation with the $m_{\nu_{\mu}}$ being determined
by $\Delta m^2_{LSND}$. It is interesting that the present LSND data has a
considerable range where the neutrinos contribute significantly to the HDM
component of the universe.

\noindent{\it \underline{Case B1.2:}}

 It is also possible to have a scenario where the
atmospheric neutrino data is fitted by the $\nu_{\mu}-\nu_s$ maximal
oscillation whereas the solar neutrino data is fitted by the
$\nu_e-\nu_{\tau}$ oscillation. Again as before, the $\Delta m^2_{LSND}$
determines the splitting between the two pairs of levels and will
determine the hot dark matter content of the universe.
 In a series of papers, Bilenky et al have 
extensively studied the tests of these models\cite{bil}.

It is possible to experimentally distinguish between these
two scenarios. The $\nu_e-\nu_s$ oscillation solution to the solar
neutrino puzzle can be tested once SNO experiment measures the neutral
current flux of the solar neutrinos since the $\nu_s$'s do not have any
neutral current interaction as opposed to the $\nu_{\mu,\tau}$. Similarly,
in the atmospheric neutrino data search for neutral current events with
pion production $\nu + N\rightarrow N+\pi^0 +N$ can discriminate between
the oscillation of the atmospheric $\nu_{\mu}$'s to $\nu_{\tau}$'s against
$\nu_s$'s\cite{vissa}. The present data appears to favor at the two
$\sigma$ level\cite{takita}
the oscillation to $\nu_{\tau}$'s but cannot be taken as a conclusive
proof.

\noindent{\it \underline{Case B1.3}:}

Another possibility is to have the three active neutrinos bunched together
with very minute mass differences such that $\nu_e-\nu_{\mu}$ oscillation
explains the solar neutrino data and the $\nu_{\mu}-\nu_{\tau}$
oscillation explains the atmospheric puzzle. The LSND data can then be
explained by including a sterile neutrino (or three of them) which is
separated in mass from
the three known neutrinos by an amount $\sqrt{\Delta m^2_{LSND}}$ and 
using indirect oscillation between the
$\nu_e-\nu_{\mu}$ via the sterile neutrino state\cite{fuller1}. It has
however been recently pointed out that this possibility may be marginally
inconsistent with the observed up-down asymmetry in atmospheric neutrino
data when combined with the CDHS and Bugey limits\cite{schwetz}.

 Let us study possible mass textures for four neutrinos.   The general
mass matrix in this case is a $4\times 4$ symmetric matrix which
has 10 nonzero entries if CP conservation is assumed. An interesting
question to explore is: what the minimum number of parameters that are
necessary to fit observations. This would then isolate useful mass
matrices with specific textures which may then lead to clues to their
theoretical origin. For this purpose let us do a bit of parameter
counting. If
the solar neutrino puzzle is assumed to be solved by the small angle MSW
mechanism,
then there must be at least five parameters in the $4\times 4$ symmetric
neutrino mass matrix: three corresponding to $\Delta m^2_{S,A,L}$ and two
small
mixings $\theta_{L}$ and $\theta_S$ (where $S,A,L$ denote respectively
the solar, atmospheric and LSND experiments). If on the other hand the
 solar neutrino puzzle is solved via the vacuum
oscillation mechanism, then one parameter is eliminated and one can make
do with four
parameters. Finally, if one adopts either the large angle MSW (between
$\nu_e$ and $\nu_s$) for solar neutrinos or the $\nu_{\mu}-\nu_s$
alternative is used to solve the atmospheric neutrino puzzle, then we have
the additional relation between parameters i.e.
\begin{eqnarray}
\frac{\Delta m^2_{S}}{\Delta m^2_{A}}\approx \frac{\Delta m^2_A}{\Delta
m^2_{L}}
\end{eqnarray}
This reduces the number of parameters to three. An example of this type
has recently been provided in Ref.\cite{rosen}. 

 Let us start with the simplest case with three parameters\cite{rosen}.
 (Here we have used the basis $(\nu_s, \nu_e, \nu_{\mu},\nu_{\tau})$.
It could be easily reshuffled to consider the $\nu_{\mu}-\nu_s$
oscillation possibility for atmospheric neutrinos.)
 \begin{eqnarray}
M=\pmatrix{m_1 &\mu& m_1-\epsilon & 0\cr
             \mu & m_1 & 0 & m_1 -\epsilon \cr
             m_1- \epsilon & 0 & m_1 & \mu\cr
             0 & m_1-\epsilon & \mu & m_1\cr}.
\end{eqnarray}
 Here $\epsilon$ is the solar mass splitting and $m_1$ is the atmospheric
mass splitting.

The next simplest case seems to involve only four
parameters and has the following form\cite{roy}:
\begin{eqnarray}
M=\pmatrix{0 &\mu_3& 0& 0\cr
             \mu_3& 0 & \epsilon & 0\cr
             0& \epsilon & 0 & m\cr
             0 & 0 & m & \delta\cr}.
\end{eqnarray}
where we choose $\delta \ll m\simeq 1$ eV. However, since in this case
the sterile neutrino has a smaller mass than the $\nu_e$ (as can be seen
by diagonalizing the above mass matrix), the solution to the solar
neutrino puzzle must involve the vacuum oscillation of the $\nu_e$ to
$\nu_s$. This can therefore be clearly tested once sufficient solar
neutrino data accumulates (say for example from Super-Kamiokande and
Borexino) in favor of or against the seasonal variation.

Going to one more parameter, we have the mass matrix 
which has one of the two following forms\cite{barger}:
\begin{eqnarray}
M=\pmatrix{\mu_1&\mu_3& 0& 0\cr
             \mu_3& 0 & 0 &\epsilon \cr
             0& 0 &\delta & m\cr
             0 & \epsilon & m &\pm\delta\cr}.
\end{eqnarray}
In this case by an appropriate choice of $\mu_1$, the sterile neutrino can
be made heavier than the elctron neutrino. As a result, one can have
useful MSW transition between the $\nu_e$ and $\nu_s$ that can help to
solve the solar neutrino puzzle via the small angle MSW solution. This
possibility has its characteristic predictions and it is expected that it
can be tested in future colliders such as the muon collider. For
discussion of CP violation in these models see Ref.\cite{new}.
\bigskip

\bigskip
\noindent{\it\underline{ B2. Six neutrino models}}
\bigskip

One obvious question that arises as we consider sterile neutrino
models is ``how many such neutrinos are there?''. Symmetry would
suggest that there are three sterile
neutrinos rather than one. In fact there are models of this type in the
literature\cite{bere,foot,chacko}. In one class of models, \cite{bere},
the additional neutrinos (i.e. the fifth and the sixth) do not play any
role in describing neutrino
observations; but in the other two\cite{foot,chacko}, they play an
essential role. The profile of
oscillation explanations in this case is very different and more symmetric
in the
sense that the both solar and atmospheric neutrino data are explained by 
maximal mixing angle vacuum oscillation between active and sterile
neutrinos (i.e. solar via $\nu_e-\nu_{es}$ and
atmospheric via $\nu_{\mu}-\nu_{\mu s}$ vacuum oscillations)\cite{bows};
the LSND results in this case is explained via generational mixing which
is small like all other charged fermion mixings. The physics of this
scheme is different from the previous ones as are the experimental tests.
We will discuss typical theories for such scenarios in subsequent
sections.

Before closing this section, it is worth pointing out that the
introduction of the sterile neutrino is such a far reaching idea that
attempts have been made to reconcile the LSND data without invoking 
oscillations in
which case solar and atmospheric neutrino puzzles as well as the LSND data
can be accomodated within the three standard neutrino framework. The idea
is
to look for consistent models where a sufficiently significant amplitude
for the anomalous decay mode of muon $\mu^+\rightarrow e^+ +\bar{\nu}_e
+\bar{\nu}_{\mu}$ exists without conflicting with other known data to
explain the $0.3\%$ rate for the observed $\bar{\nu}_e$'s.
However it has been pointed out\cite{peter}  that
the muonium to anti-muonium ($M-\bar{M}$) transition is
related to the anomalous muon decay in several models (such as
left-right models or supersymmetric models with R-parity breaking) in such
a way that 
the existing experimental bounds from PSI\cite{psi} on  $M-\bar{M}$
transition suppresses the anomalous muon decay amplitude far below that
required to explain the LSND data. Thus barring some really drastic new
idea to explain the LSND data as an anomalous muon decay, the confirmation
of the LSND data in future would require the existence of additional one
or more sterile neutrinos\footnote{There have also been attempts to
explain the atmospheric neutrino data by non-oscillation
mechanisms\cite{nonosc} such as flavor changing interactions or neutrino
decays etc. In a recent paper, Bergman et al\cite{bgp} have argued that
the first
alternative runs into problem with large flavor changing effects in $\tau$
decays, which have not been observed and therefore that mechanism is 
theoretically not very plausible.}. 
It will then be dependent on the data as to
which of the scenarios with extra sterile neutrinos will win in the end.

\section{Why neutrino mass necessarily means physics beyond the standard
model ?}

As is wellknown, the standard model is based on the gauge group
$SU(3)_c\times SU(2)_L\times U(1)_Y$ group under which the quarks and
leptons transform as described in the Table I.

The electroweak symmetry $SU(2)_L\times U(1)_Y$ is broken by the vacuum
expectation of the Higgs doublet $<H^0>=v_{wk}\simeq 246$ GeV, which gives
mass to the gauge bosons and the fermions, all fermions except the
neutrino. Thus the neutrino is massless in the standard model, at the tree
level. There are several questions that arise at this stage. What happens
when one goes beyond the above simple tree level approximation ? Secondly,
do nonperturbative effects change this tree level result ? Finally,  how
will this result be modified when the quantum gravity effects
are included ?

\begin{center}
{\bf Table I}
\end{center}

\begin{center}
\begin{tabular}{|c||c|}
\hline\hline
 Field &  gauge  transformation \\ \hline\hline
 Quarks $Q_L$ & $(3,2, {1\over 3})$\\
 Righthanded up quarks $u_R$ &  $(3, 1, {4\over 3})$ \\
Righthanded down quarks  $ d_R$ &  $(3, 1,-\frac{2}{3})$\\
Lefthanded  Leptons $L$ & $(1, 2 -1)$ \\
 Righthanded leptons  $e_R$ & $(1,1,-2)$ \\
Higgs Boson $\bf H$ & $(1, 2, +1)$ \\
Color Gauge Fields  $G_a$ & $(8, 1, 0)$ \\
Weak Gauge Fields  $W^{\pm}$, $Z$, $\gamma$ & $(1,3+1,0)$ \\
\hline\hline
\end{tabular}
\end{center}

\noindent {\bf Table caption:} The assignment of particles to the standard
model gauge group $SU(3)_c\times SU(2)_L\times U(1)_Y$.

\bigskip

The first and second questions are easily answered by using the B-L
symmetry of the standard model. The point is that since the standard model
has no $SU(2)_L$ singlet neutrino-like field, the only possible mass terms
that are allowed by Lorentz invariance are of the form
$\nu^T_{iL}C^{-1}\nu_{jL}$, where $i,j$ stand for the generation index and
$C$ is the Lorentz charge conjugation matrix. Since the $\nu_{iL}$ is part
of the $SU(2)_L$ doublet field and has lepton number +1, the above
neutrino mass term transforms as an $SU(2)_L$ triplet and furthermore, it
violates total lepton number (defined as $L\equiv L_e+L_{\mu}+L_{\tau}$)
by two units. However, a quick look at the standard model Lagrangian
convinces one that the model has exact lepton number symmetry both
before and after symmetry breaking; therefore such terms can never arise
in perturbation theory.
Thus to all orders in perturbation theory, the neutrinos are massless.
As far as the nonperturbative effects go, the only known source is the
weak instanton effects. Such effects could effect the result if they 
broke the lepton number symmetry. One way to see if such breaking
occurs is to look for anomalies in lepton number current conservation from
triangle diagrams. Indeed $\partial_{\mu}j^{\mu}_{\ell}= c W \tilde{W} +
c' B\tilde{B}$ due to the contribution of the leptons to the triangle
involving the lepton number current and $W$'s or $B$'s. Luckily, it turns
out that the anomaly contribution to the baryon number current
nonconservation has also an identical form, so that the $B-L$ current 
$j^{\mu}_{B-L}$ is conserved to all orders in the gauge couplings. As a
consequence, nonperturbative effects from the gauge sector cannot induce
$B-L$ violation. Since the neutrino mass operator described above violates 
also $B-L$, this proves that neutrino masses remain zero even in the
presence of nonperturbative effects.

Let us now turn to the effect of gravity. Clearly as long as we treat
gravity in perturbation theory, the above symmetry arguments hold since
all gravity coupling respect $B-L$ symmetry. However, once nonperturbative
gravitational effects e.g black holes and worm holes are
included\cite{giddings}, there is no guarantee that global symmetries will
be respected in the low energy theory. The intuitive way to appreciate the
argument is to note that throwing baryons into a black hole does not lead
to any detectable consequence except thru a net change in the baryon
number of the universe. Since one can throw in an arbitrary numnber of
baryons into the black hole, an arbitrary information loss about the net
number of missing baryons would prevent us from defining a baryon
number of the visible
universe- thus baryon number in the presence of a black hole can not be an
exact symmetry. Similar arguments can be made for any global charge such
as lepton number in the standard model. A field theoretic parameterization 
of this statement is that the effective low energy Lagrangian for the
standard model in the presence of black holes and worm holes etc must
contain baryon and lepton number violating terms. In the context of the
standard model, the only such terms that one can construct are
nonrenormalizable terms of the form $~LH LH/M_{P\ell}$. After gauge
symmetry breaking, they lead to neutrino masses. It however becomes
immediately clear that these masses are not enough to understand present
observations since they are at most of order $~v^2_{wk}/M_{P\ell}\simeq
10^{-5}$ eV\cite{akm} and
as we discussed in the previous section, in order to solve the atmospheric
neutrino problem, one needs masses at least three orders of magnitude
higher.

Thus one must seek physics beyond the standard model to explain observed
evidences for neutrino masses.

\section{Scenarios for small neutrino mass without right handed neutrinos}

There exist a number of extensions of the standard model that lead to
neutrino masses. Even restricting ourselves to the cases where the
neutrinos are Majorana particles there come to mind at least three
different mechanisms for neutrino masses. All these models have lepton
number violation built into them and therefore lead to a plethora of new
phenomenological tests. The existence or nonexistence of the right handed
neutrinos divides these models into two broad classes: one class that
uses right handed neutrinos and another that does not. In this section we
consider models that do not introduce right handed neutrinos to understand
small neutrino masses.

 As discussed in
section II, the electrical neutrality of the neutrino allows for the
existence of two types of mass terms consistent with Lorentz invariance:  
the Majorana mass, which violates lepton number
but does not require the inclusion of a right handed neutrino and the
Dirac (or combined Dirac-Majorana) mass term which requires the existence
of a right handed neutrino. In this subsection we consider models where 
the neutrino Majorana mass arises without the right handed neutrino. The
argument of section IV then implies that there must be
violation of the B-L quantum number either by interactions or by the
vacuum state. This leads to two classes of models which
we discuss below.

\subsection{ Radiative generation of neutrino masses} 

There are two classes of models where introduction of lepton number
violating interactions leads to radiative generation of small neutrino
masses. One is the Zee model\cite{zee} and the second one is the Babu
model\cite{babu}. Let is first discuss the Zee model. In this case, one
adds a charged $SU(2)_L$-singlet field $\eta^+$ to the standard model
along with a second Higgs doublet $H'$. This allows the following
additional Yukawa coulings beyond those present in the standard
model:
\begin{eqnarray}
{\cal L}_Y(\eta) = \Sigma_{\alpha \beta} f_{\alpha\beta}L^T_{\alpha} 
C^{-1}L_{\beta} \eta^+ +~h.c.
\end{eqnarray}
Note that the coupling $f$ is antisymmetric in the family indices $\alpha$
and $\beta$. Due to the presence of the extra Higgs field, there are also
additional terms in the Higgs potential but the one of interest in
connection with the neutrino masses is $\eta^* H^T\tau_2 H'$. It is then
easy to see that while the neutrino masses vanish at the tree level they
arise from one loop diagrams due to the exchange of $\eta$ fields in
combination with the second Higgs (see Fig. 1). The typical strength of
these diagrams is of order $\frac{m^2_{\ell}f}{16\pi^2 M}$\cite{lincoln}.
The dominant
contribution to the $\nu_{e,\mu}$ mass in these models comes from the
$\tau$ lepton in the loop and can be estimated to be of order 1-10 keV for
$f\simeq
10^{-2}$ and the $\eta H H'$ coupling also of order $10^{-2}$-$10^{-1}$.
These values for the masses are much bigger than being contemplated in the
context of neutrino puzzles. Thus in the opinion of this author, Zee model
cannot account for the present observations without further ado.

Let us now pass to the second class of models\cite{babu}. In this case
one adds to
the standard model the fields $\eta^+$ as before and a doubly charged
field $h^{++}$ but not a second Higgs doublet. The new Yukawa couplings of
the model then are given by (repeated indices are summed):
\begin{eqnarray}
{\cal L}_Y(\eta, h) = f_{\alpha\beta} \eta L_{\alpha} L_{\beta}
+f'_{\alpha\beta} h^{++} e^T_{\alpha, R}C^{-1}e_{\beta, R} + ~ h.c.
\end{eqnarray}
The Higgs potential has the following term that is of interest in
connection with the neutrino masses i.e. $\lambda''v_{wk} h^*\eta\eta $.
The model leads to nonzero contribution to neutrino masses at the two
loop level via the diagrams of figure 2.

The typical estimate for the neutrino
mass in this case is $\sim \frac{f^2f' m^2_{\ell}
\lambda''}{(16\pi^2)^2 M}$.
For $f\sim 10^{-1.5}\sim f'\sim \lambda''$, the two loop contribution to
neutrino mass is of order $0.01$ eV which is of the right order of
magnitude for
solving the present neutrino puzzles. No excessive fine tuning of
couplings is needed. In building realistic models, one has to however pay
attention to possible flavor changing neutral current effects such as the
$\mu\rightarrow e\gamma$ decay for which there exist rather stringent
constraints from the recent MEGA experiment\cite{MEGA} : $B(\mu\rightarrow
e+\gamma)\leq 10^{-11}$. This will put constraints on the parameters as
follows:
$f_{13}f_{23}\leq 10^{-5}$ and $(f'_{13}f'_{23} + f'_{12}f'_{11,22})\leq
10^{-5}$. We do not get into detailed model building for this case.

One process that distinguishes the second class of models from the first
is the muonium to antimuonium ($M-\bar{M}$) oscillation which is mediated
at the tree level by the doubly charged bosons\cite{chang}. The present
limit on  the strength of the effective four-Fermi interaction
describing $M-\bar{M}$ transition is given by $G_{M-\bar{M}}\leq 3\times
10^{-3}G_F$\cite{psi}. For a 100 GeV $h^{++}$ boson, this implies that
$h_{ee}h_{\mu\mu}\leq 3\times 10^{-4}$, which is a nontrivial constraint
on the model. Since for the neutrino masses to be in the interesting
range, the Higgs masses should not be more than a TeV, further improvement
of the $M-\bar{M}$ transition limit can provide important information on
this model.

\subsection{High mass Higgs triplet and induced neutrino masses}

Another way to generate nonzero neutrino masses 
without using the righthanded neutrino is to include in the standard model 
an $SU(2)_L$ triplet Higgs field with $Y=2$ so that the electric charge
profile of the members of the multiplet is given as follows:
$(\Delta^{++}, \Delta^{+}, \Delta^0)$.  This allows an additional Yukawa
coupling of the form $ f_LL^T\tau_2{\bf \tau}L.{\bf \Delta}$, where the
$\Delta^0$ couples to the neutrinos. Clearly $\Delta$ field has $L=2$. 
When $\Delta^0$ field has a
nonzero vev, it breaks lepton number by two units and leads to Majorana
mass for
the neutrinos. There are two questions that arise now: one, how does the
vev arise in a model and how does one understand the smallness of the
neutrino masses in this scheme. There are two answers to the first
question: One can maintain exact lepton number symmetry in the model and
generate the vev of the triplet field via the usual ``mexican hat''
potential. There are two problems with this case. This leads to the
massless triplet Majoron\cite{gel} which has been ruled out by LEP data on
Z-width. Though it is now redundant it may be worth pointing out that in
this model smallness of the neutrino mass is not naturally understood.

There is however another way to generate the induced vev keeping a large
but positive mass ($M_{\Delta}$) for the triplet Higgs boson and allowing
for a lepton number violating coupling $M\Delta^* H H$. In this case,
minimization of the potential induces a vev for the $\Delta^0$ field when
 the doublet field acquires a vev: 
\begin{eqnarray}
v_T\equiv <\Delta^0>= \frac{M v^2_{wk}}{M^2_{\Delta}}
\end{eqnarray}
Since the mass of the $\Delta$ field is invariant under $SU(2)_L\times
U(1)_Y$, it can be very large connected perhaps with some new scale of
physics. If we assume that $M_{\Delta}\sim M\sim 10^{13}$ GeV or so, we
get $v_T\sim$ eV. Now in the Yukawa coupling $ f_LL^T\tau_2{\bf
\tau}L.{\bf \Delta}$, since the $\Delta^0$ couples to the neutrinos,
its vev
leads to a neutrino mass in the eV range or less depending on the value of
the Yukawa couplings\cite{many}. We will see later when we discuss the
seesaw models that unlike those models, the neutrino mass in this case is
not hierarchically dependent on the charged fermion masses. Another
point worth emphasizing is that unlike the previous radiative
scenarios, this model is
more in the spirit of grand unification and can in fact be implemented in
models\cite{book} such as those based on the SU(5) group where there is no
natural place for the right handed neutrino.

\subsection{Baryogenesis problem in models without right handed
neutrinos}

While strictly the models just discussed provide a way to understand the
small neutrino masses without fine tuning, they may face problems in
explaining the origin of matter in the universe. Let us first consider the
triplet vev models. It was shown by Ma and Sarkar\cite{many}, that the
decay of triplet Higgs to leptons provides a way to generate enough
baryons in the model. However for that to happen, one must satisfy one of
Sakharov's three condition that the decay particle which leads to baryon
or lepton asymmetry must be out of equilibrium. This requires that
when the temperature of the universe equals the decaying particle's mass,
its decay rate must be smaller than the expansion rate of the universe
i.e.
\begin{eqnarray}
\frac{f^2_L M_{\Delta}}{12\pi}< \sqrt{g^*}\frac{M^2_{\Delta}}{M_{P\ell}}
\end{eqnarray}
This implies a lower limit on the mass $M_{\Delta}\geq \frac{f^2_L
M_{P\ell}}{12\pi \sqrt{g^*}}$. For $f_L\sim 10^{-1}$ as would be required
by the atmospheric neutrino data, one gets conservatively, $M_{\Delta}\geq
10^{13}$ GeV. The problem with such a large mass arises from the fact
that in an inflationary model of the universe, the typical reheating
temperature dictated by the gravitino problem of supergravity is at most
$10^9$ GeV. Thus there is an inherent conflict\cite{del} between the
standard inflationary picture of the universe and the baryogenesis in the
simple triplet model for neutrino masses.

As far as the radiative models go, they have explicit lepton violating
terms
in the Lagrangian which are significant enough to keep these interactions
in equilibrium for all interesting low temperatures ($T<10^{12}$ GeV or
so). So it is not clear how one would ever generate any baryon number in
this class of models.

\section{Seesaw mechanism and left right symmetric unification models for
small neutrino masses}

A very natural and elegant way to generate neutrino masses is to include
the right
handed neutrinos in the standard model. However the inclusion of the right
handed neutrinos transforms the dynamics of gauge models in a profound
way. To clarify what we mean, note that in the standard model (that does
not contain a $\nu_R$) the $B-L$ symmetry is only linearly anomaly free
i.e. $Tr[(B-L)Q^2_a]=0$ where $Q_a$ are the gauge generators of the
standard model but $Tr(B-L)^3\neq 0$. This means that $B-L$ is only a
global symmetry and cannot be gauged. However as soon as the $\nu_R$ is
added to the standard model, one gets $Tr[(B-L)^3]=0$ implying that the
B-L symmetry is now gaugeable and one could choose the gauge group of 
nature to be either $SU(2)_L\times U(1)_{I_{3R}}\times U(1)_{B-L}$ or
$SU(2)_L\times SU(2)_R\times U(1)_{B-L}$, the latter being the gauge group
of the left-right symmetric models\cite{moh}. Furthermore the presence of
the $\nu_R$ makes the model quark lepton symmetric and leads to a
Gell-Mann-Nishijima like formula for the elctric charges\cite{marshak}
i.e.
\begin{eqnarray}
Q= I_{3L}+I_{3R}+\frac{B-L}{2}
\end{eqnarray}
The advantage of this formula over the charge formula in the standard
model is that in this case all entries have a physical
meaning. It also provides a natural understanding of Majorana nature of
neutrinos
as can be seen by looking at the distance scale where the $SU(2)_L\times
U(1)_Y$ symmetry is valid but the left-right gauge group is broken. In
that case, one gets
\begin{eqnarray}
\Delta Q=0= \Delta I_{3L}:\\ \nonumber
\Delta I_{3R}~=~-\Delta \frac{B-L}{2}
\end{eqnarray}
We see that if the Higgs fields that break the left-right gauge group
carry righthanded isospin of one, one must have $|\Delta L| = 2$ which
means that the neutrino mass must be Majorana type and the theory will
break lepton number by two units.

Let us now proceed to discuss the left-right symmetric model and
demonstrate how the seesaw mechanism emerges in this model.

The gauge group of the theory is
SU$(2)_L \, \times$ SU$(2)_R \, \times$ U$(1)_{B-L}$ with quarks and
leptons transforming as doublets under SU$(2)_{L,R}$.
In Table 2, we present transformation properties of the quark, lepton and
Higgs fields of the model under the gauge group.
~~~~~~~~~~
\begin{center}
{\bf Table II}
\end{center}

\begin{center}
\begin{tabular}{|c|c|} \hline\hline
Fields           & SU$(2)_L \, \times$ SU$(2)_R \, \times$ U$(1)_{B-L}$ \\
                 & representation \\ \hline
$Q_L$                & (2,1,$+ {1 \over 3}$) \\
$Q_R$            & (1,2,$ {1 \over 3}$) \\
$L_L$                & (2,1,$- 1$) \\
$L_R$            & (1,2,$- 1$) \\
$\phi$     & (2,2,0) \\
$\Delta_L$         & (3,1,+ 2) \\
$\Delta_R$       & (1,3,+ 2) \\
\hline\hline
\end{tabular}
\end{center}

\noindent{\bf Table caption} Assignment of the fermion and Higgs
fields to the representation of the left-right symmetry group.

\bigskip

The first task is to specify how the left-right symmetry group breaks to
the standard model i.e. how one
breaks the $SU(2)_R\times U(1)_{B-L}$ symmetry so that the successes of
the standard model
including the observed predominant V-A structure of weak interactions at
low energies is reproduced. Another question of naturalness that also
arises simultaneously is that since the charged fermions and the
neutrinos are treated completely symmetrically (quark-lepton symmetry)
in this model, how does one understand the smallness of the neutrino
masses compared to that of the other fermions.

It turns out that both the above problems of the LR model have a common
solution. The process of spontaneous breaking of the $SU(2)_R$ symmetry
that suppresses the V+A
currents at low energies also solves the problem of ultralight neutrino
masses. To see this let us write the Higgs fields in terms of its
components:
\begin{eqnarray}
\Delta~=~\left(\begin{array}{cc} \Delta^+/\sqrt{2} & \Delta^{++}\\
\Delta^0 & -\Delta^+/\sqrt{2} \end{array}\right); ~~
\phi~=~\left(\begin{array}{cc} \phi^0_1 & \phi^+_2\\
\phi^-_1 & \phi^0_2 \end{array}\right)
\end{eqnarray}
 All these
Higgs fields have Yukawa couplings to the fermions given symbolically as
below.
\begin{eqnarray}
{\cal L_Y}= h_1 \bar{L}_L\phi L_R +h_2\bar{L}_L\tilde{\phi}L_R\nonumber \\
+ h'_1\bar{Q}_L\phi Q_R +h_2'\bar{Q}_L\tilde{\phi}Q_R
\nonumber\\
+f(L_LL_L\Delta_L +L_RL_R\Delta_R) +~ h.c. \end{eqnarray}
The $SU(2)_R\times U(1)_{B-L}$ is broken down to the standard model
hypercharge $U(1)_Y$ by choosing $<\Delta^0_R>=v_R\neq 0$ since this
carries
both $SU(2)_R$ and $U(1)_{B-L}$ quantum numbers. It gives mass to the
charged and neutral righthanded gauge bosons i.e. $M_{W_R}= gv_R$ and
$M_{Z'}=\sqrt{2} gv_R cos\theta_W/\sqrt{cos 2\theta_W}$. Thus by
adjusting the value of $v_R$ one can suppress the right handed current
effects in both neutral and charged current interactions arbitrarily
leading to an effective near maximal left-handed form for the charged
current weak interactions at low energies.

The fact that at the same time the neutrino masses also become small can
be
seen by looking at the form of the Yukawa couplings. Note that the f-term
leads to a mass for the right handed neutrinos only at the scale $v_R$.
Next as we break the standard model symmetry by turning on the vev's for
the $\phi$ fields as $Diag<\phi>=(\kappa, \kappa')$, we not only
give masses to the $W_L$ and the $Z$ bosons but also to the quarks and the
leptons. In the neutrino sector the above Yukawa couplings after
$SU(2)_L$ breaking by $<\phi>\neq 0$ lead to the Dirac masses
for the neutrino
connecting the left and right handed neutrinos. In the two component
neutrino language, this leads to the following mass matrix for the
$\nu, N$ (where we have denoted the left handed neutrino by $\nu$ and the
right handed component by $N$).
\begin{eqnarray}
M=\left(\begin{array}{cc} 0 & h\kappa \\
h\kappa & fv_R\end{array}\right)
\end{eqnarray}
By diagonalizing this $2\times 2$ matrix, we get the light neutrino 
eigenvalue to be $m_{\nu}\simeq \frac{(h\kappa)^2}{fv_R}$ and the heavy
one to be $fv_R$. Note that typical
charged fermion masses are given by $h'\kappa$ etc. So since $v_R\gg
\kappa, \kappa'$, the light neutrino mass is automatically suppressed.
This way of suppressing the neutrino masses is called the seesaw mechanism
\cite{grsyms}. Thus in one stroke, one explains the smallness of the
neutrino mass as well as the suppression of the V+A currents\footnote{
There is an alternative class of left-right symmetric models where small
neutrino masses can arise from radiative corrections, if one chooses only
doublet Higgses to break the gauge symmetry and heavy vector-like
charged fermions to understand fermion mass hierarchies. In these models
the neutrinos are Dirac particles\cite{gab}.}.

In deriving the above seesaw formula for neutrino masses, it has been
assumed that the vev of the lefthanded triplet is zero so that the
$\nu_L\nu_L$ entry of the neutrino mass matrix is zero. However, in most
explicit models such as the left-right model which provide an explicit
derivation of this formula, there is an induced vev for the $\Delta^0_L$
of order $<\Delta^0_L> = v_T\simeq \frac{v^2_{wk}}{v_R}$. In the presence
of this term the seesaw formula undergoes a fundamental change. Let us
therefore distinguish between two types of seesaw formulae:

\bigskip
\noindent {\bf \underline{Type I seesaw formula}}
\bigskip

\begin{eqnarray}
M_{\nu}\simeq - M^T_{D}M^{-1}_{N_{R}}M_D
\end{eqnarray}
where $M_D$ is the Dirac neutrino mass matrix and $M_{N_R}\equiv fv_R$ is
the right handed neutrino mass matrix in terms of the $\Delta$ Yukawa
coupling matrix $f$.

\bigskip
 \noindent{\bf \underline{Type II seesaw formula}}
\bigskip

\begin{eqnarray}
M_{\nu}\simeq f\frac{v^2_{wk}}{v_R} - M^T_{D}M^{-1}_{N_{R}}M_D
\end{eqnarray}

Note that in the type I seesaw formula, what appears is the square of the
Dira neutrino mass matrix which in general expected to have the same
hierarchical structure as the corresponding charged frermion mass matrix.
In fact in some specific GUT models such as SO(10), $M_D=M_u$. This is the
origin of the common statement that neutrino masses given by the seesaw 
formula are hierarchical
i.e.
$m_{\nu_e}\ll m_{\nu_{\mu}}\ll m_{\nu_{\tau}}$ and even a more model
dependent statement that $m_{\nu_e} : m_{\nu_{\mu}} : m_{\nu_{\tau}}=
m^2_u : m^2_c : m^2_t$.

On the other hand if one uses the type II seesaw formula, there is no
reason to expect a hierarchy and in fact if the neutrino masses turn out
to be degenerate as discussed before as one possibility, one way
to understand this may be to use the type II seesaw formula.

\bigskip
\noindent{\bf \underline{Type III seesaw formula}}

\bigskip

While the above seesaw formulae are extremely helpful in building models
with ultralight neutrino masses, there are circumstances where they are
not helpful. An obvious example is one where the righthanded neutrino mass 
matrix is singular due to additional symmetries (e.g.
$L_e-L_{\mu}-L_{\tau}$ say), in which case its inverse
does not exist and alternative ways to understand lightness of neutrinos
must be explored. There is one such mechanism in literature which we will
call type III seesaw which involves a $3\times 3$ seesaw
pattern\cite{valle} rather than the just described $2\times 2$ one. 
For one generation, it involves three fermions $(\nu, N, S)$, where
$\nu$ and $N$ are the usual left and right handed neutrinos and
$S$ is a singlet neutrino (singlet under the left-right group). The
relevant seesaw matrix is given by 
\begin{eqnarray}
{\cal M}_{\nu}=\left(\begin{array}{ccc}
0 & m_D & 0 \\
m_D & 0 & M\\
0 & M & \mu \\ \end{array}\right)
\end{eqnarray}
We assume that $\mu, m_D \ll M$ in which case the three eigenvalues are
given by
\begin{eqnarray}
m_{\nu}\sim \frac{m^2_D\mu}{M^2}
\end{eqnarray}
Its generalization to the case when each element is a matrix is obvious.
The important point here is to note that the matrix $\mu$ is in the
numerator. As  a result, if we want the light neutrino mass matrix to be
singular (say $L_e-L_{\mu}-L_{\tau}$ invariant), then this can
be built into the matrix $\mu$ and one can then use the seesaw formula
given here to understand the smallness of the neutrino masses. In the
context of left-right models this kind of a structure
for neutrino masses arises if the triplets above are replaced by $B-L=1$
doublets and we supplement the model with one singlet fermion per
generation. ({\it Show this by explicit
construction and work out the example with $L_e-L_{\mu}-L_{\tau}$
symmetry}). Another feature of the type III seesaw is that one could
choose the B-L breaking scale to be much lower than in the case of the
types I nd II.

Let us now address the question: to what extent one can understand the
details of the neutrino masses and mixings using the seesaw formulae.
 The answer to this question is quite model dependent. While 
one can say that there exist many models which fit the observations, none 
(except a few) are completely predictive. The problem in general is that  
the seesaw formula of type I has 12
parameters which is why its predictive power is so limited. One number
that is predicted
in a class of seesaw models based on the SO(10) group that embodies the
left-right symmetric unification model or the SU(4)-color is the tau
neutrino mass. In this
class of models, one maintains the quark lepton symmetry in the leading
order so that one has the relation $m_t(M_U)=m^D_{\nu_{\tau}}$. The tau
neutrino mass is then given by the seesaw formula to be $m_{\nu_{\tau}}
\simeq \frac{m^2_t}{M_{N_{3R}}}$. In one class of string inspired models,
where the $SU(2)_R$ symmetry and the GUT symmetry break at the same scale,
the right handed neutrino masses are generated by nonrenormalizable
operators and they are given in the simplest approximation by
$\frac{fv^2_R}{M_{P\ell}}\simeq f10^{14}$ GeV. If $f=1$, then one gets
$m_{\nu_{\tau}}\simeq 0.03$ eV which is the right value to fit the
atmospheric neutrino data. The rest of the data such as the maximal
mixing with the muon neutrino etc are not predicted without further
assumptions. Even the prediction of the $\nu_{\tau}$ mass requires an
assumption that the Yukawa coupling $f$ must be unity.

\subsection{ SO(10) realization of the seesaw mechanism}

The most natural grand unified theory for the seesaw mechanism is the
SO(10) model. In this paragraph, some of the salient features of this
realization are summarized. From the previous paragraph, we learn that the
simplest left
right model with B-L=2 triplets provides provides a direct realization of
the seesaw mechanism. In the context of the SO(10) model, the first point
to note is that the {\bf 16}-dimensional spinor representation contains
all the fermions of each generation in the standard plus the right handed
neutrino. Thus the right handed neutrino is automatic in the SO(10) model.
Secondly, in order to break the B-L symmetry present in the SO(10) group,
one may use either the Higgs multiplets in {\bf 16} or {\bf 126}
dimensional representation. Under the left-right symmetric group
$SU(2)_L\times SU(2)_R\times U(1)_{B-L}\times SU(3)_c$, these fields
decompose as follows:
\begin{eqnarray}
{\bf 16}_H= (2, 1, 1/3, 3)\oplus (1,2, -1/3, 3^*)\oplus (2,1, -1,1)\oplus
(1,2, +1, 1)\\ \nonumber
{\bf 126}= (1, 1, -2/3, 3)\oplus (1,1, 2/3, 3^*)\oplus (3, 1, -2, 1)\oplus
(1,3, +2, 1) + ..... 
\end{eqnarray}
where the ellipses denote other multiplets that have no role in neutrino
mass discussion. In order to break the B-L symmetry it is the last entry
in each of the multiplets whose neutral elements need to pick up a large
vev. Note however
that the first one (i.e. the ${\bf 16}_H$) does not have any
renormalizable coupling with the {\bf
16} spinors which contain the $\nu_R$ whereas there is a renormalizable
SO(10) invariant coupling of the form ${\bf 16}{\bf 16}{\bf \bar{126}}$.
for the second multiplet. Therefore if we decided to stay with the
renormalizable model, then one
would need a {\bf 126} dimensional representation to implement the seesaw
mechanism whereas if we used the {\bf 16} to break the B-L symmtery, we
would require nonrenormalizable couplings of the form ${\bf 16}^2 {\bf
\bar{16}_H}^2/M_{P\ell}$. This has important implications for the B-L
scale.
In the former case, the B-L breaking scale is at an intermediate level
such as $\sim 10^{13}$ GeV or so whereas in the latter case, we can have
B-L scale coincide with the GUT scale of $2\times 10^{16}$ GeV as in the
typical SUSYGUT models\cite{tasi}.

Another advantage of SO(10) models in understanding neutrino masses is
that if one uses only the {\bf 10} dimensional representation for giving
masses to the quarks and leptons, one has the up
quark mass matrix $M_u$ being equal to the Dirac mass matrix of the
neutrinos which goes into the seesaw formula. As a result, if we work in a
basis where the up quark masses are diagonal so that all CKM mixings come
from the down mass matrix, then the number of arbitrary parameters in the
seesaw formula goes dowm from 12 to 6. Thus even though one cannot
predict neutrino masses and mixings, the parameters of
the theory get fixed by their values as inputs. This may then be testable
thru its other predictions. In this
model however, there are tree level mass relations in the down sector such
as $\frac{m_d}{m_s}=\frac{m_e}{m_{\mu}}$ which are renormalization group
invariant and are in disagreement with observations. It may be possible in
supersymmetric models to generate enough one loop corrections out of the
supersymmetry breaking terms (nonuniversal) to save the situation.

There is one very special cases where all 12 of these parameters are
predicted by the
quark and lepton masses\cite{babum}. The Higgs content of this model is
only one {\bf 10} and one {\bf 126}. Their Yukawa couplings involve
only nine parameters all of which along with input vevs of the doublets 
in both these multiplets are fixed by the quark masses, mixings and the 
three charged lepton masses. The important point is that the same {\bf
126} responsible for the fermion masses also has a vev along the
$\nu_R\nu_R$ directions so that it generates the right handed neutrino
mass matrix. Thus in the light neutrino sector it is a completely
parameter free model. But this minimal SO(10) model cannot fit
both the solar and the atmospheric neutrino data and is therefore ruled
out. Recently other SO(10) models have been
considered where under different assumptions, the atmospheric and solar
neutrino data can be explained together\cite{many1}. These models have
many interesting features, which we do not go into here for lack of space.

In the minimal supersymmetric left-right model, an analogous situation 
happens where the neutrino Dirac masses are found to be equal to the
charged lepton masses\cite{dutta}. Thus in this model too, one has only
six parameters to describe the neutrino sector and once the neutrino data
is fitted all parameters in the model are fixed so that one has
predictions that can be tested. For instance, it has been emphasized in
Ref.\cite{dutta} that there is a prediction for the $B(\tau\rightarrow
\mu+\gamma)$ in this model that is about two orders of magnitude below the
present limits\cite{cleo} and could therefore ne used to test the model.

There is yet another class of models where by assigning $U(1)$ charges to
the fermions of the standard model as well as the $\nu_R$ fields\cite{u1},
one restricts the Dirac as well as Majorana mass matrices (for the
$\nu_R$). One then assumes that the $U(1)$ charges originate from strings
or some high scale physics.

Finally let us comment that in models where the light neutrino mass is
understood via the seesaw mechanism that uses heavy righthanded neutrinos,
there is a very simple mechanism for the generation of baryon asymmetry of
the universe. Since the righthanded neutrino has a high mass, it decays at
a high temperature to generate a lepton asymmetry\cite{fuku} and this
lepton asymmetry is converted to baryon asymmetry via the sphaleron
effects\cite{kuzmin} at lower temperature. It also turns out that one of
the necessary conditions for sufficient leptogenesis is that the right
handed neutrinos must be heavy as is required by the seesaw mechanism. To
see this note that one of Sakharov conditions for leptogenesis is that the
right handed neutrino decay must be slower than the expansion rate of the
universe at the temperature $T\sim M_{N_R}$. The corresponding condition
is:
\begin{eqnarray}
\frac{h^2_{\ell}M_{N_R}}{16\pi}\leq \sqrt{g^*}\frac{M^2_{N_R}}{M_{P\ell}}
\end{eqnarray}
This implies that $M_{N_R}\geq \frac{h^2_{\ell}M_{P\ell}}{16\pi
\sqrt{g^*}}$. For the second generation, it implies that
$M_{N_{2R}}\geq 10^{13}$ GeV and for the third generation a value even
higher. Note that these are above the inflation reheating upper bound
alluded to before. However for the first generation, it is about $10^{8}$
GeV so that there is no conflict with the gravitino bound on the reheating
temperature. Incidentally, the leptogenesis condition also imposes limits
on the matrix elements of the right handed neutrino mass, thereby reducing
the arbitrariness of the seesaw predictions slightly.

\section{Naturalness of degenerate neutrinos}

In this section we like to discuss some issues related to the degenerate
neutrino hypothesis. Recall that this is the only scenario that fits all
observations if one does not include LSND (say it is not confirmed by
MiniBooNE) and the universe has 10\% to 20\% of its matter in the form of
neutrinos. Thus it is appropriate to discuss how such models can arise in
theoretical schemes and how stable they are under radiative corrections.

The first point already alluded to before and first made in \cite{cald1}
is that degenerate neutrinos arise naturally in models that employ the
type
II seesaw since the first term in the mass formula is not connected to the
charged fermion masses. One way that has been discussed is to consider
schemes where one uses symmetries such as SO(3) or SU(2) or permutation
symmetry $S_4$\cite{deg} so that the Majorana Yukawa couplings $f_i$ are
all equal. This then leads to the dominant contribution to all
neutrinos being equal. This symmetry however must be broken in the charged
fermion sector in order to explain the observed quark and lepton masses.
Such models consistent
with known data have been constructed based on SO(10) as well as other
groups. The interesting point about the SO(10) realization is that the
dominant contributions to the $\Delta m^2$'s in this model comes from the
second term in the type II seesaw formula which in simple models is
hierarchical. It is of course known that if the MSW solution to the solar
neutrino puzzle is the right solution (or an energy independent solution),
then we have $\Delta m^2_{solar} \ll \Delta m^2_{ATMOS}$. In fact if we
use the fact true in SO(10) models that $M_u=M_D$, then we have $\Delta
m^2_{ATMOS}\simeq m_0\frac{m^2_t}{fv_R}$ and $\Delta
m^2_{SOLAR}\simeq m_0\frac{m^2_c}{fv_R}$ where $m_0$ is the common mass
for the three neutrinos. It is interesting that for $m_0\sim $ few eV and
$fv_R\approx 10^{15}$ GeV, both the $\Delta m^2$'s are close to the
required values.

An interesting theoretical issue about these models has been raised in
several recent papers\cite{deg2}. It has been noted that even though one
may have tree level models with a degenerate neutrino spectrum, it is not
clear that this mass degeneracy will survive the radiative corrections. In
fact it has been
convincingly argued that in models with maximal mixings the departure from
degeneracy may be significant. This provides a further challenge to model
building and one must ensure that should maximal mixing models win the
day, the degeneracy is preserved upto at least to two loop.

\section{Theoretical understanding of the sterile neutrino}

If the existence of the sterile neutrino becomes
confirmed say, by a confirmation of the LSND observation of
$\nu_{\mu}-\nu_e$
oscillation or directly by SNO neutral current data to come in the   
early part of the next century, a key theoretical challenge will be to
construct an underlying theory that embeds the
sterile neutrino along with the others with appropriate mixing pattern,
while naturally explaining its ultralightness.

If a sterile neutrino was introduced into the standard 
model, the gauge symmetry would not forbid its bare mass, implying that
there would be no reason for it mass to be small. It is a common
experience in
physics that if a particle has mass lighter than normally expected on the
basis of known symmetries, then it is an indication for the existence of
new
symmetries. This line of reasoning has been pursued in recent literature
to understand the ultralightness of the sterile neutrino by using new
symmetries beyond the standard model.

There are several suggestions for this new symmetry that might help us to
accomodate an ultralight $\nu_s$ and in the process lead us to new
classes of extensions of physics beyond the standard model. We will
consider two examples (i) one based on the $E_6$ grand unification model
and (ii)another based on the possible existence of mirror matter in
the universe. There are also other theoretical models for the sterile
neutrino that we involve different assumptions\cite{smir} and we do not
discuss them here.

A completely different way to understand the ultralightness of the sterile 
neutrino is to introduce large extra dimensions in a Kaluza-Klein
framework\cite{dimo} and not rely on any symmetries. In these models, the
sterile neutrino is supposed to live in the bulk; therefore if its
Lagrangian mass is zero, then the mass of the first Kaluza-Klaein mode is
inversely proprotional to the size of the extra dimension. Thus the extra
dimension size of order of a millimeter would lead to sterile neutrino
masses
in the range of milli eV's which are of interest for solar neutrino
oscillations. There have been recent attempts\cite{lore} to build
realistic models using a slight variation of this idea\cite{satya}. This
class of models are distinguished from others in their prediction of
infinite tower of sterile neutrinos, which at the moment seems perfectly
consistent with observations. Eventually, this feature may provide a way 
to test such theories. We do not elaborate on these models any
further.

\section{$E_6$ model for the sterile neutrino}

 $E_6$\cite{gursey} is an interesting and viable
unification group beyond the SO(10) group and as such has been extensively
discussed in literature\cite{e6}. In this model, matter belongs to the
{\bf 27} dimensional
representation of the $E_6$ group which under its $SO(10)\times U(1)$  
subgroup 
decomposes to ${\bf 16}_{+1}\oplus {\bf 10}_{-2}
\oplus {\bf 1}_{+4}$  (the subscripts represent the U(1) charges). The
{\bf 16} is well known to contain the left and the right handed neutrinos
(to be denoted by us as $\nu_i$ and $\nu^c_i$, $i$ being the family
index). The {\bf 10} contains two neutral colorless fermions which behave
like neutrinos but are $SU(2)_L$ doublets and the last neutral colorless 
fermion in the {\bf 27}, which we identify as the sterile neutrino is the
one contained in the SO(10) singlet multiplet (denoted by $\nu_{is}$).
This has all the properties desired of a sterile neutrino.
Thus in this model there
are three sterile neutrinos which will be denoted by the corresponding
flavor label and we will show first how the small mass for both the active
and the sterile neutrino results from a generalization of the the seesaw
mechanism as a
consequence of the symmetries of the group\cite{ma}.  Furthermore, we
will see how as a consequence of the smallness
of the Yukawa couplings of the standard model, we will not only get
maximal mixing between the active and the sterile neutrinos of each   
generation but also the necessary ultra-small $\Delta m^2$ needed in the
vacuum oscillation solution to the solar neutrino puzzle without fine
tuning of
parameters\cite{chacko}. Thus in this model, both the solar and the
atmospheric neutrino puzzles are solved by the maximal vacuum oscillation
of active to sterile neutrinos.

In general in this
model, we will have for each generation a $5\times 5$ ``neutrino'' mass
matrix. To give an essence of the basic steps that lead to the seesaw
mechanism, it is necessary to describe the
symmetry breaking of $E_6$. We work with a supersymmetric $E_6$ and use
three pairs of
${\bf 27}+\bar{\bf 27}$ representations and one {\bf 78}-dim. field
for symmetry breaking. The pattern of symmetry breaking is as follows:

1) $<27_1>$ and $<\overline{27}_1>$ have GUT scale vevs in the SO(10)
   singlet direction.

2) $<27_{16}>$ and $<\overline{27}_{16}>$ have GUT scale vevs in the
   $\nu$ and $\nu^c$ directions respectively. They break SO(10) down
   to SU(5).

3) The $<78_{[1,45]}>$ completes the breaking of SU(5) to the standard
model gauge group at the GUT scale. We assume the VEVs reside both in the
   adjoint and in the singlet of SO(10).

4) $<27_{10}>$ and $<\overline{27}_{10}>$ contain the Higgs doublets of
the MSSM. It is assumed that $H_u$ and $H_d$ are both linear combinations
 arising partially from the $<27_{10}>$ and partially from the     
 $<\overline{27}_{10}>$

In addition to the above there is another field labelled by $27^\prime$
whose $\nu^c$ component mixes with a singlet S and one linear combination
of this pair 
remains light below the GUT scale. As a consequence of
radiative symmetry breaking  this picks up a VEV at the electroweak
scale. This was shown in Ref.\cite{chacko}. The remaining components
of $27^\prime$ have GUT scale mass.
Let us now write down the relevant terms in the superpotential that lead
to a $5\times 5$ ``neutrino'' mass matrix of the form we desire.
 To keep matters simple let us
ignore generation mixings, which can be incorporated very trivially. 
   
\begin{eqnarray}
W~=~\lambda_i \psi_i\psi_i 27_{10}  + f_i \psi_i\psi_i 27^\prime +  
\frac{\alpha_i}{M_{P\ell}}\psi_i\psi_i 27_{1} 78_{[1,45]}
+\frac{\gamma_i}{M_{P\ell}}\psi_i\psi_i \overline{27}_{16}
\overline{27}_{16}
\end{eqnarray}
We have shown only a subset of allowed terms in the theory that play a
role in the neutrino mass physics and believe
that it is reasonable to assume a discrete symmetry (perhaps in the   
context of a string model) that would allow only this subset.
In any case since we are dealing with a supersymmetric theory, radiative
corrections will not generate any new terms in the superpotential.
   
Note that in Eq. (41), since it is the first term that leads to lepton and
quark masses of
various generations, it carries a generation label and obeys a
hierarchical pattern, whereas the $f_i$'s not being connected to known
fermion masses need not obey a hierarchical pattern. We will from now on
assume that each $f_i\approx 1$.

After substituting the vevs for the Higgs fields in the above equation,
we find a $5\times 5$ mass matrix  of the following form for the
neutral lepton fields of each generation in the basis $(\nu, \nu_s, \nu^c,
E^0_u, E^0_d)$:
\begin{eqnarray}
M~=~ \left(\begin{array}{ccccc}
0 & 0 & \lambda_i v_{u} & f_i v' & 0 \\
0 & 0 & 0 & \lambda_i v_{d} & \lambda_i v_{u} \\
\lambda_i v_{u} & 0 & M_{\nu^c,i} & 0 & 0 \\
f_i v' & \lambda_i v_{d} & 0 & 0 & M_{10,i} \\
0 & \lambda_i v_{u} & 0 & M_{10,i} & 0 \end{array} \right)
\end{eqnarray}
Here $M_{\nu^c,i}$ is the mass of the right handed neutrino and $M_{10,i}$   
is the mass of the entire {\bf 10}-plet in the {\bf 27} matter multiplet.
Since {\bf 10} contains two full SU(5) multiplets, gauge coupling
unification will not be effected even though we choose its mass to be
below the GUT scale.     

Note that the $ 3\times 3$ mass matrix involving the $(\nu^c, E^0_u,
E^0_d)$ have superheavy entries and will therefore decouple at low
energies. Their effects on the spectrum of the light neutrinos will be
dictated by the seesaw mechanism\cite{grsyms}. The light neutrino mass   
matrix involving $\nu_i, \nu_{is}$ can be written down as:
\begin{eqnarray}
M_{light}\simeq \frac{1}{M_{\nu^c,i}} \left(\begin{array}{ccc}
 \lambda_i v_{u} & f_i v' & 0 \\
 0 & \lambda_i v_{d} & \lambda_i v_{u}\end{array}\right)   
\left(\begin{array}{ccc}
 1 & 0 & 0 \\
 0 & 0 & \epsilon \\
 0 & \epsilon & 0 \end{array} \right)\left(\begin{array}{cc}
\lambda_i v_{u} & 0 \\
f_i v' & \lambda_i v_{d}  \\
0 & \lambda_i v_{u}  \end{array} \right)
\end{eqnarray}  

where $\epsilon_i = M_{10,i}/ M_{\nu^c,i}$. Note that $\epsilon_i $ is
expected to
be of order one. This leads to the $2\times 2$ mass matrix for the $(\nu,
\nu_c)$ fields of each generation the form,

\begin{eqnarray}
M_i= m_{0i}\left(\begin{array}{cc}
\lambda^2_i & \lambda_i \bar{f}_i\\
\lambda_i \bar{f}_i & \lambda^2_i \bar{\epsilon_i}\end{array}\right)
\end{eqnarray}

Here $ m_{0i} = \frac{v_{u}^2}{M_{\nu^c,i}}$, $\bar{f}_i = f_i\epsilon_i
v'/v_u$,
and $ \bar{\epsilon_i} = 2 \epsilon_i cot{\beta}$. Taking $M_{Pl} \sim
10^{19}GeV$, $M_{GUT} \sim 10^{16}$ and reasonable values of the unknown 
parameters e.g. $\alpha_i\approx
0.1$, $\gamma_i\approx 0.1$,
$f_i\approx 1$, $v'\approx v_u$, we get
$m_{0i}\simeq 20$ eV and $\epsilon\approx 1$ which leads us
to the desired pattern of masses and mass differences where active
neutrinos of each generation mix with the corresponding sterile neutrino
maximally and the $\Delta m^2$'s scale as the cube of the 
corresponding charged fermion mass. Thus if we fix the $\Delta
m^2_{ATMOS}\sim 10^{-3}$ eV$^2$, then we get the vacuum oscillation
solution for the solar neutrino puzzle.               

This model also provides an explanation of the LSND results and the
smallness of the mixing angle observed in the experiment is now similar to
that observed in the quark sector and therefore easily understood by the
same mechanisms that provide an explanation of the small quark mixing
angles.   

\section{Mirror universe model of the sterile neutrino}

The second suggestion that explains the ultralightness of the
$\nu_s$ is the mirror matter model\cite{bere,foot} where the basic idea 
is that there is a complete duplication of matter and forces in the
universe (i.e. two sectors in the universe with matter and gauge forces
identical prior to symmetry breaking).
The mirror sector of the model will then have three light neutrinos
which will not couple to the Z-boson and would not therefore have been 
seen at LEP. We denote the fields in the mirror sector by a prime over
the standard model fields. We will call the $\nu'_i$ as the sterile
neutrinos of which  
we now have three. The lightness of $\nu'_i$ is dictated by the mirror   
$B'-L'$ symmetry in a manner parallel to what happens in the standard
model. Thus the ultralightness of the sterile neutrinos is understood in
the most ``painless'' way.

The two ``universes'' are assumed to communicate only via
gravity\cite{okun,blinni} or other forces that are
equally weak. This leads to a mixing between the neutrinos of the two
universes and can cause oscillations between $\nu_e$ of our
universe to $\nu'_e$ of the parallel one in order to explain for example
the solar neutrino deficit.

  At an overall level, such a picture emerges quite naturally in
superstring theories which lead to $E_8\times E_8^\prime$ gauge theories
below the Planck scale with both $E_8$s connected by gravity.
For instance, one may assume the sub-Planck GUT
group to be a subgroup of $E_8\times E^\prime_8$ in anticipation of
possible future string embedding. One may also imagine the visible
sector and the mirror sector as being in two different D-branes, which    
are then necessarily connected very weakly due to exchange of massive bulk  
Kaluza-Klein excitations.

In the mirror model, both sectors can either remain identical after
symmetry breaking or there can be asymmetry. We will consider the second
scenario. This was suggested in Ref.\cite{bere}.

As suggested in Ref.\cite{bere}, we will assume that the process of
spontaneous symmetry breaking introduces asymmetry between the two
universes
e.g. the weak scale $v^\prime_{wk}$ in the
mirror universe is larger than the weak scale $v_{wk}= 246$ GeV in our
universe. The ratio of the two weak scales $\frac{v'_{wk}}{v_{wk}}\equiv
\zeta$ is the only parameter that enters the fit to the solar
neutrino data. It
was shown in Ref.\cite{bere} that with $\zeta\simeq 20-30$, the   
gravitationally generated neutrino masses\cite{akm} can provide a
resolution of the solar neutrino puzzle
(i.e. one parameter generates both the required $\Delta m^2_{e-s}$ and the
mixing angle $sin^22\theta_{e-s}\simeq 10^{-2}$). There can also be a
large angle MSW fit with reasonable choice of parameters e.g. smaller
$\zeta$ but with coefficients of higher dimensional operators allowed to
vary between 0.3 to 3.

There are other ways to connect the visible sector with the mirror sector
using for instance a bilinear term involving the righthanded neutrinos
from the mirror and the visible sector.  An $SO(10)\times SO(10)$
realization of this idea was studied in detail in
a recent paper\cite{brahma}, where a complete realistic model
for known particles and forces including a fit to the fermion masses and    
mixings was worked out and the resulting predictions for the masses and
mixings
for the normal and mirror neutrinos were presented. In this model,
the fermions of each generation are assigned to the ${\bf (16, 1)\oplus
(1, 16^\prime)}$ representation of the gauge group. The $SO(10)$
symmetry is broken down to the left-right symmetric model by the
combination
of ${\bf 45\oplus 54}$ representations in each sector. The $SU(2)_R\times
U(1)_{B-L}$ gauge symmetry in turn is broken by the ${\bf 126 \oplus
\overline {126}}$ representations. These latter fields serve two purposes:
first, they guarantee automatic
R-parity conservation and second, they lead to the see-saw
suppression for the neutrino masses. We do not get into detailed
discussion of the masses and mixings of neutrinos in such a model here.
The mixing between the two sectors is caused by a multiplet belonging to
the representation ${\bf (16, 
16^\prime)+(\overline{16},\overline{16}^\prime)}$. If
the mass of this last multiplet is kept at the GUT scale, the mixing
between the two righthanded neutrinos caused by this is large and the
ensuing effects on the light neutrino mixings are small.              

Since the mirror matter model has many ultralight particles, the issue of
consistency with big bang nucleosynthesis must be addressed.
Recall that present observations of Helium and deuterium
abundance can allow for at most $4.53$ neutrino species\cite{bbn} if the
baryon to photon ratio is chosen appropriately. Since the model has three
extra neutrinos, an extra photon and the extra $e^+e^-$ pair, apriori, the
effective neutrino count in the model could be as alrge as 8.2. However,
in the asymmetric mirror model, since the neutrinos decouple
above 200 MeV or so due to weakness of the mirror Fermi coupling, their
contribution at the time of nucleosynthesis
is negligible (i.e. they contribute about $0.3$ to $\delta N_{\nu}$.)
On the other hand the mirror photon could be completely in equilibrium at
$T=1$ MeV so that it will contribute $\delta N_{\nu}=1.11$. There is also
a contribution from the mirror electron-positron pair of $N_{\nu}=2$. All
together the total contribution to $N_{\nu}$ is less than $6.4$. So to
solve this problem, we invoke an idea called asymmetric inflation whereby
the reheating temperature in the mirror sector is lower than the reheating
temperature in the familiar sector. Models with this kind of possibility 
was discussed in \cite{dolgov}. 
   
There may be another potentially very interesting application of the idea
of the mirror universe. It appears that
there may be a crisis in understanding the microlensing
observations\cite{micro}.  It has
to do with the fact that the best fit mass for the 14 observed
microlensing events by the MACHO and the EROS group is around $0.5
M_{\odot}$
and it appears difficult to use normal baryonic objects of similar mass  
such as white dwarfs to explain these events, since they
lead to a number of cosmological and astrophysical problems\cite{freese}.   
Speculations have been advanced tha this crisis may also be resolved
by the postulate that the MACHOs with $0.5 M_{\odot}$ may be mirror   
stars\cite{bere1,teplitz} which would then have none of the difficulties
that arise from interpretations in terms of conventional baryonic matter.
In this model, mirror baryons can form the dark matter of the universe so
that there is no need for a neutralino dark matter.

\section{Conclusions and Outlook}

In this review, we have tried to provide a brief look at the new physics
implied by the discovery of neutrino oscillations. We start with a  
summary of the possible
textures for neutrino masses implied by the present data with and without
the inclusion of the LSND data. Several three, four and six neutrino
scenarios have been
outlined. We then briefly describe the various ideas primarily aimed at
understanding the smallness of the neutrino masses both in the context of
grand or simple unified models. Finally two theories that can naturally
incorporate an ultralight sterile neutrino are discussed. Clearly, there
is considerable subjective judgement used in the selection of models and
an apology is due to all authors whose models are not discussed here.
Limitations of space is certainly one simple excuse that can be given. In
any case, the final story in this field is yet to be written and it could 
very easily be that the models described here do not eventually win out.
Even that will be a very valuable piece of information since the ideas
touched on here are certainly some of the more salient ones around now. 
But it is the fervent hope of
this author (presumably shared by many workers in the field) that some of
the ideas described in the literature (and reviewed here) will survive the
test of time. The ``ball'' is right now in the court of the
experimentalists and we pin our hopes on the large number of continuing 
(SNO, Borexino, K2K, MINOS) as well as planned (KAMLAND, ICARUS, LENS,
GENIUS, CUORE, ORLAND etc) 
experiments in this field. On the theoretical side, true progress can be
said to have been made only in understanding the smallness of the neutrino
masses in different scenarios but the complete picture of mixing angles
is far from being at hand, although there are many models that under
various assumptions lead to realistic fits. One of the hardest theoretical
problems is to understand the ultrasmall mass difference that would be
required if the vacuum oscillation solution to the solar
neutrino problem eventually wins the race (although there exist a handful
of interesting suggestions\cite{chacko,ma3,foot}). One will then have to
check whether the radiative corrections in these models destabilize the
result.

The bottom line is that the field of neutrino
mass has become one of the most vibrant and exciting fields in particle
physics as we move into the new millenium and is at the moment the only
 beacon of new physics beyond the standard model.

\bigskip

\noindent{\bf Acknowledgement}

This work was supported by the National Science Foundation grant number
PHY-9802551. I would like to acknowledge many useful conversations on the
subject of this review with K. S. Babu, D. Caldwell, Z. Berezhiani, B.
Kayser, S. Nussinov and V. L. Teplitz.

\begin{figure}[htb] \begin{center}
\epsfxsize=7.5cm
\epsfysize=7.5cm
\mbox{\hskip -1.0in}\epsfbox{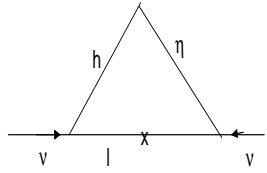}
\caption{One loop diagram for neutrino mass in the Zee model
\label{Fig.1}}
\end{center}
\end{figure}

\begin{figure}[htb] \begin{center}
\epsfxsize=7.5cm
\epsfysize=7.5cm
\mbox{\hskip -1.0in}\epsfbox{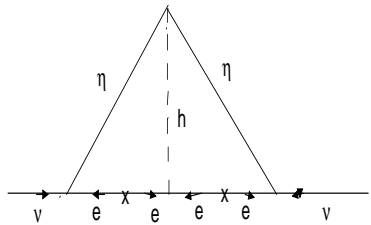}
\caption{Two loop diagram for neutrino mass in the Babu model
\label{Fig.2}}
\end{center}
\end{figure}

\end{document}